**Fitting the Low Energy Spectra of Cosmic Ray Primary Nuclei from**

**C to Fe as Measured on Voyager 1:  The Need to Modify the Leaky Box**

**Exponential Distribution of Path Lengths to Explain the Data**

**and the Resulting Distribution of Cosmic Ray Sources**

**W.R. Webber**

New Mexico State University, Astronomy Department, Las Cruces, New Mexico, 88003, USA

**ABSTRACT**

The intensities of the low energy part of the spectra of primary cosmic ray nuclei including C, O, Ne, Mg, Si and Fe measured by Voyager 1 beyond the heliopause are deficient relative to the spectra measured at energies above ~100 MeV/nuc as calculated using a standard Leaky Box Model with the path length a function of rigidity. Modifications to the normal exponential distribution of path lengths at a single rigidity as is used in a simple LBM will provide a good fit to this new Voyager 1 data at low energies. These modifications, sometimes called a truncation, lead to a deficiency of short path lengths relative to an exponential distribution. This deficiency in the intensities can be described by a truncation parameter = 0.12. This modification can be produced in several ways including a non-uniform local distribution of cosmic ray sources. A uniform source distribution in the galactic plane that is deficient in sources within 0.2-0.4 Kpc of the Sun is indicated by the data. Further studies of these low energy spectra in more detail will improve these estimates and help define other features of the local galactic distribution of these cosmic rays.

**Introduction**

In a previous paper (Webber, 2015) we have discussed specific features of the interstellar spectra of the primary (source) nuclei of H through Fe in the energy range above and below ~100 MeV/nuc that are measured on Voyager 1 beyond the heliopause (Cummings, et al., 2015). Above ~100 MeV/nuc the spectra of all of these nuclei have been shown to be consistent with simple rigidity spectra $\sim P^{-2.26}$ extending up to ~100 GeV/nuc and above. Below 100 MeV/nuc the C-Fe source nuclei spectra diverge strongly and in a specific way from the predictions using a simple Leaky Box Model (LBM) in that the measured intensities of all nuclei, C, O, Ne, Mg, Si and Fe, are much less than predicted by an extension of these higher rigidity spectra to lower rigidities (Webber, 2015). The H and He nuclei observed by Voyager do not show a similar deficiency of low energy particles (e.g., Webber and Higbie, 2015) and will be discussed in a separate paper.

In the LBM model used here, which considers finite but unstructured propagation and which leads to an exponential path length distribution (Cowsik, et al., 1967) as well as in models with additional structure beyond the Leaky Box such as the GALPROP models (Strong and Moskalenko, 1998), a key concept in investigating the differences between these and other models is the path length distribution itself, e.g., the possible evidence for a deficit of short path lengths or truncation as it is called (Shapiro and Silberberg, 1970; Garcia-Munoz, et al., 1987; Webber, 1993). This deficit has been used in these papers to investigate the deviation of the cosmic ray sources from a locally uniform distribution, for example. Numerically this deficit of short path lengths has been described as follows (e.g., Sharpiro and Silberberg, 1970). For a path length $\lambda_2 = f \cdot \lambda_1$ where $\lambda_1$ is the normal path length, a convenient representation of the modified non-exponential path length distribution is

$$P(x) = (exp^{-x/\lambda}_1 - exp^{-x/f\lambda}_1) \qquad (1)$$

where f is a "truncation" parameter (<1.0).

The mean path length $\lambda$, in g/cm$^2$ traversed by the primary nuclei has a maximum ~10 g/cm$^2$ at ~120 MeV/nuc, decreasing at both lower and higher energies as determined by new Voyager 1 and AMS-2 data on the B/C ratio (Webber and Villa, 2016). This variation in the path length can be accounted for by changes in the rigidity dependence of the diffusion coefficient assuming input source rigidity spectra that are the same exponent for all charges and which is independent of rigidity for all charges (Webber, 2015).

In this paper we will show that the low energy spectral differences that are observed for source nuclei between C and Fe at V1 can be explained in terms of a deficit of short path lengths in the normal exponential path length distribution itself and this deficit may then be used to provide an insight into the distribution of the sources of cosmic rays in the plane of the galaxy on a relatively local scale.

## **The Observations**

In this paper we present in Figure 1 the intensities of C+O, Ne+Mg+Si and Fe nuclei measured by Voyager (Cummings, et al., 2015).  In this Figure we also show the LBM calculations of the spectra of these nuclei.  The turnover in the spectra of these nuclei at low energies relative to the higher energy measured spectra and the calculation is evident for all nuclei.  To illustrate this deficiency in intensity in more detail we next compare the intensities at 100 MeV/nuc, both measured and calculated, with the intensities at other energies.  This energy acts as a fulcrum point between the high and low energy spectra and helps to illustrate the differences between the low energy spectra of the different nuclei.  The original intensity references are found in Cummings, et al., 2015 and used in Webber, 2015.

Figure 2 shows these intensity ratios normalized at 100 MeV/nuc for (C+O)/2 nuclei. Figure 3 shows the ratios normalized at 100 MeV/nuc for the sum Ne+Mg+Si nuclei and Figure 4 for Fe nuclei.  In each figure the data points are shown in red and the predictions (in black) are for a simple LBM with $\lambda$=22.3 $\beta$ $P^{-0.47}$ and a value of $P_0$=1.0 GV where the path length dependence becomes $\lambda$=33.7 $\beta^{3/2}$ at lower rigidities.  The data shows intensities below ~100 MeV/nuc that are much reduced below those predicted from this standard LBM.  Above ~100 MeV/nuc, however, the data and predictions of the LBM are consistent with essentially the same LIS spectra for all of these nuclei which become ~$P^{-2.26}$ above 10 GeV/nuc.  This dependence extends in energy up to and above 100 GeV/nuc as described by Webber, 2015.

To explain these low energy spectra we have tried various alternatives to the standard LBM propagation parameters, including larger values of $P_0$ and a different dependences of $\lambda$ below $P_0$. We have also increased the ionization loss parameter. None of these variations reproduces well the V1 observations at low energies.

To obtain spectra similar to those observed thus appears to require a modification to the path length distribution itself. This modification is taken to be that illustrated by equation (1) where the key parameter is the truncation parameter. This approach has been used in previous studies to interpret spectral and intensity differences of various cosmic ray nuclei (Shapiro and Silberberg, 1970; Garcia, Munoz, et al., 1987)

In Figures 2, 3 and 4 the calculations for various truncation parameters which are equal to 0.06, 0.12 and 0.20 are shown in blue. These modifications to the path length distribution introduce a deficiency of short path lengths relative to a normal exponential PLD as indicated in Figure 5 where a mean path length = 4 g/cm$^2$ exponential distribution is used as an example for a pure LBM. The absence of short path lengths, P(x), for x < 1 g/cm$^2$ relative to an exponential distribution is obvious in these examples in Figure 5.

These modifications of the path length distribution give intensities below 100 MeV/nuc that bracket the new Voyager observations for all 3 groups of nuclei using a truncation parameter ~0.12. This value, in fact, provides an excellent fit to the most accurately measured C+O spectra.

Note that in Figures 2, 3 and 4 we also show the latest GALPROP calculations of the various spectra as obtained from Cummings, et al., 2016.

**Discussion**

The exponential distribution of path lengths in a LBM was originally discussed by Davis, 1960. Owens (1976) rigorously calculated the path length distribution in a LBM at relativistic energies and found that the exponential path length distribution was well satisfied in agreement with a simple LBM. At smaller path lengths the exponential distribution expected in a LBM could be influenced by specific propagation conditions, boundary conditions and the source distribution. Some of these are, for example, shown by the GALPROP calculations in Figures 2, 3 and 4, which exhibit intensities that are between a pure LBM and one modified by spatial asymmetries.

Modifications to the path length distribution similar to those described above in this paper using a truncation parameter have been considered earlier (Shapiro and Silberberg, 1970; Garcia-Munoz, et al., 1987). A double or nested LBM, where some of the material traversed by cosmic rays is located near the sources, has also been considered (Cowsik and Wilson, 1973; Cowsik, Burch and Madziwa-Nussinov, 2014). Specific models for the distribution of cosmic ray sources, leading to a truncated PLD have also been proposed (Lezniak and Webber, 1979; Higden and Lingenfelter, 2003), for example.

These early examples of the modifications to the simple LBM were applied to explain the cosmic ray spectra observed by balloon and spacecraft borne detectors in the 1970's and 1980's without much success. One reason for the lack of success in these earlier studies was that corrections to the spectra for the large solar modulation effects at low energies were just too formidable. So this field has languished in the state described by Webber, 1993, for over 20 years until Voyager 1 crossed the heliopause into a region where the solar modulation was essentially zero (Webber and McDonald, 2013). One exception to this was the GALPROP code which first made its appearance in Strong and Moskalenko, 1998. Later an on-line version of the program appeared (Vladimirov, A. E., et al. 2011). This program introduced considerable structure in the galaxy. This structure is an important aspect of the propagation of galactic cosmic rays from their source regions. Newly available gamma-ray maps of the galaxy from a few MeV to a few GeV from Fermi-Lat (Ackermann, et al., 2011) could then be interpreted using this program in terms of the distribution of energetic cosmic ray nucleons and electrons that produced gamma rays of comparable energies in the galaxy.

The passage of V1 beyond the heliopause has opened a new window, the study of the spectral shape of the individual cosmic ray primary nuclei at energies below ~100 MeV/nuc. Our application of the LB diffusion model to a description of these spectra indicates that the spectra of the primary nuclei C→Fe nuclei are deficient in low energy particles relative to the same set of parameters as applied in the same LBM used to describe the diffusive propagation of the higher energy C→Fe nuclei. For some reason the path length distribution at each rigidity no longer appears to be an exponential at these low energies. This modified path length distribution has been described, for calculation purposes here, in terms of a sum of two exponentials where one is smaller than the other by a truncation factor, f. From these calculations we have determined that, for a value of f=0.12, the deficiency in the primary nuclei C → Fe intensities below ~100 MeV/nuc that is observed at Voyager can be explained by a deficiency in the path length distributions from the exponential distribution as shown in Figure 5.

There are several possibilities to explain the origin of such a modified path length distribution, one of which is a non-uniformity of the cosmic ray source distribution when viewed from the Sun. We will concentrate here on this possibility, because there are at least two papers that actually provide calculations that can then be used to determine the shape of modified path length distributions for possible source distributions. In one of these papers (Lezniak and Webber, 1979), a distribution of sources in the plane of the galaxy is taken to be non-uniform on a scale size, $\ell$, near the Sun. The path length distributions that are obtained in this paper for $\ell$/L=0.15 and 0.25, where L is an assumed galactic plane half-thickness, are also shown in Figure 5. These distributions are similar to those obtained using a truncation parameter = 0.12 also shown in Figure 5. Therefore for a disk thickness, L ~1 Kpc, this could imply a non-uniformity scale of the source distribution ~0.2-0.4 Kpc; e.g., there are no nearby cosmic ray sources within this distance over the time scale for cosmic ray confinement in the galaxy which has been determined to be ~15 M yr. (Yanasak, et., al, 2001).

Another model considers the spatial distribution of O-B sources which are candidates for the origin of cosmic rays (Higdon and Lingenfelter, 2003), in a myriad source model. The path length distributions are shown in this paper for several possibilities for the diffusion coefficient and the assumed thickness of the disk. If we choose the one where the diffusion coefficient = 0.3 pc and the disk thickness = 1.5 Kpc, the PLD obtained from Figure 3 of their paper is shown as a blue curve in Figure 5.

## Summary and Conclusions

In a previous paper we pointed out a significant intensity deficit in the low energy spectra relative to the spectra above ~100 MeV/nuc for the primary cosmic ray C→Fe nuclei as observed beyond the heliopause by V1 (Webber, 2015).  Above 100 MeV/nuc the interstellar spectra of these nuclei are all consistent within $\pm10\%$ between 100 MeV/nuc and 100 GeV/nuc with source spectra that are ~$P^{-2.26}$.  The corresponding spectra below ~100 MeV/nuc are not fit well using reasonable variations of the parameters used in a simple LBM which has an exponential distribution of path lengths.

In this paper we have found that modified versions of the LBM with a truncated path length distribution at path lengths less than a few g/cm$^2$ give an excellent fit to these primary spectra down to the lowest measured energies of about 10 MeV/nuc.  These modified PLD can be caused by several effects related to both the propagation and to the source distribution. Herein we make use of two sets of calculations which provide possible PLD from assumed "local" cosmic ray source distributions (Lezniak and Webber, 1979; Higdon and Lingenfelter, 2003).  Both models are for the local source distribution within a disk shaped galaxy of thickness 1-2 Kpc.  They lead to modified PLD which look similar to the calculations for the example calculation made in this paper which uses a truncation parameter =0.12.  This would indicate that the nearest "source" of cosmic rays within the last 1.5-10$^7$ year is at least ~0.2-0.4 Kpc from the Sun.

The study of the low energy spectra of these and other primary cosmic rays on Voyager thus opens a new window for understanding the origin and acceleration of cosmic rays.  The Voyager observations show that the distribution of galactic cosmic rays is not completely unstructured and uniform, which is the case for a simple LBM, but contains features, particularly at low energies, that can be used to examine this structure on a local (Kpc) scale.  This work and that of GALPROP are a start in understanding how productive this new window will be.

**Acknowledgements:**  The author appreciates the efforts of all of the Voyager CRS team, Ed Stone, Alan Cummings, Nand Lal and Bryant Heikkila.  It is a pleasure to work with them as we try to understand and interpret the galactic cosmic ray spectra that had previously been hidden from us by the solar modulation. We also thank JPL for their continuing support of this program for over 40 years. This work would not have been possible without the assistance of Tina Villa.

# FIGURE CAPTIONS

**Figure 1:**    Spectra of (C+O)/2, Ne+Mg+Si, and Fe nuclei measured at V1 beyond the heliopause.  The statistical errors on the individual C+O points are $\pm$ 3%, Ne+Mg+Si points are $\pm$ 5% and Fe points are $\pm$ 10%.  Propagated spectra using a LBM for these nuclei between 3-1000 MeV/nuc are shown as solid black lines.  These propagated spectra fit the Voyager data above ~100 MeV/nuc and other data up to 100 GeV/nuc and beyond as described in Webber, 2015.  They also fit the H and He spectra measured by Voyager (Webber and Higbie, 2015).

**Figure 2:**  C and O intensities measured at Voyager 1 at low energies, normalized to a value of 1.0 at 100 MeV/nuc (in red).  The normalized intensities at 100 MeV/nuc are shown in the figure in units of $P/(m^2 \cdot sr \cdot s \cdot MeV/nuc)$.  The propagated spectra using a value of $P_0=1.0$ in our standard LBM are shown in black.  The modified LBM spectra with truncation parameters = 0.06, 0.12 and 0.20 are shown in blue.  A predicted spectrum using the GALPROP model (Cummings, et al., 2016) is shown as a dashed blue line.

**Figure 3:**  Same as Figure 2 but for the sum of Ne, Mg and Si nuclei measured by V1 and predicted by the various models.

**Figure 4:**  Same as Figure 2 but for Fe nuclei measured by V1 and predicted by the various models.

**Figure 5:**  The path length distribution at small values of the mean path length X, for an exponential distribution with X = 4 $g/cm^2$ (red line) and this exponential distribution modified using truncation parameters = 0.06, 0.12 and 0.20 (black lines) as described in equation 1.  The path length distributions obtained in the Lezniak and Webber, 1979, and Higdon and Lingenfelter, 2003, models as described in the text are also shown as orange and blue lines respectively.

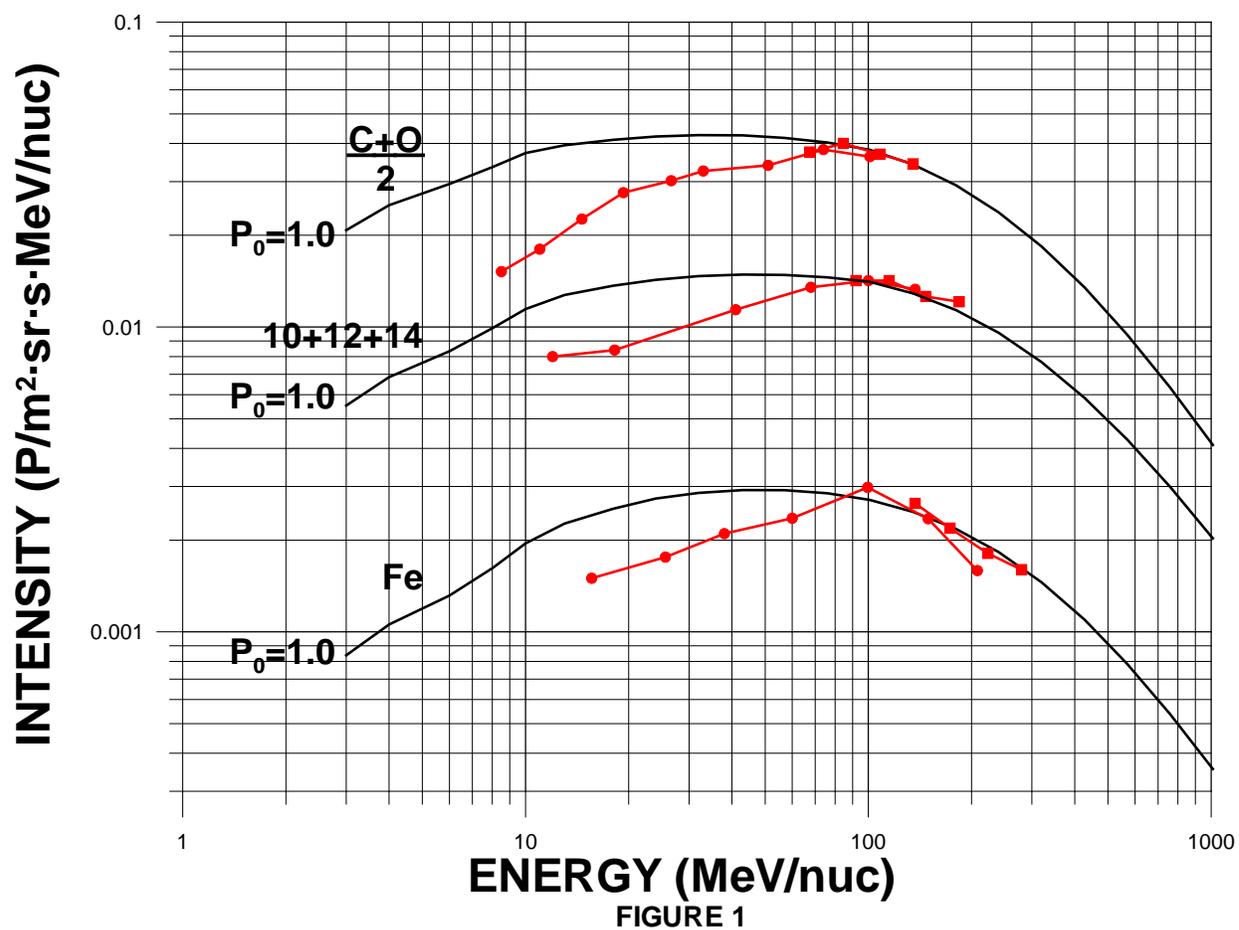

FIGURE 1

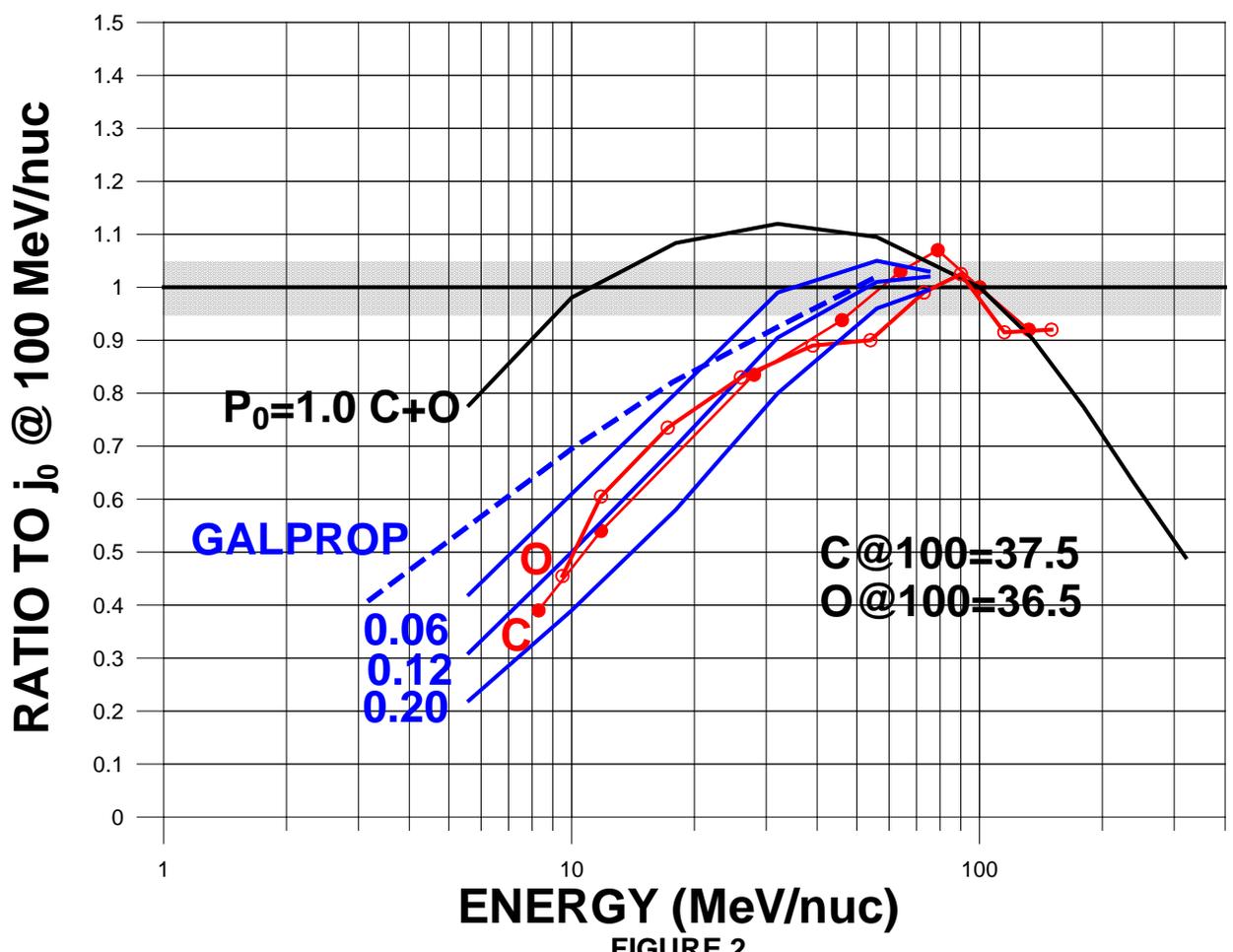

FIGURE 2

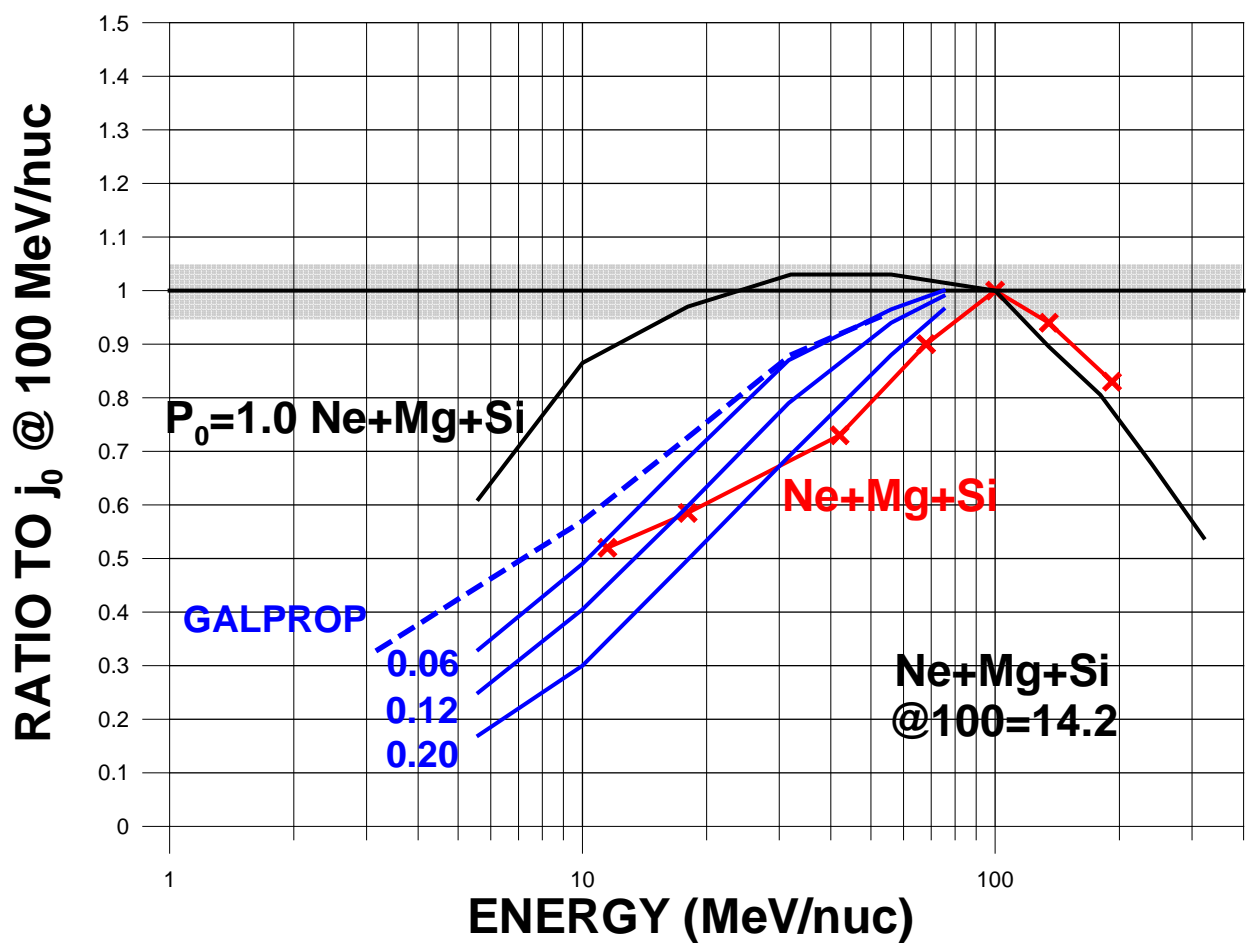

FIGURE 3

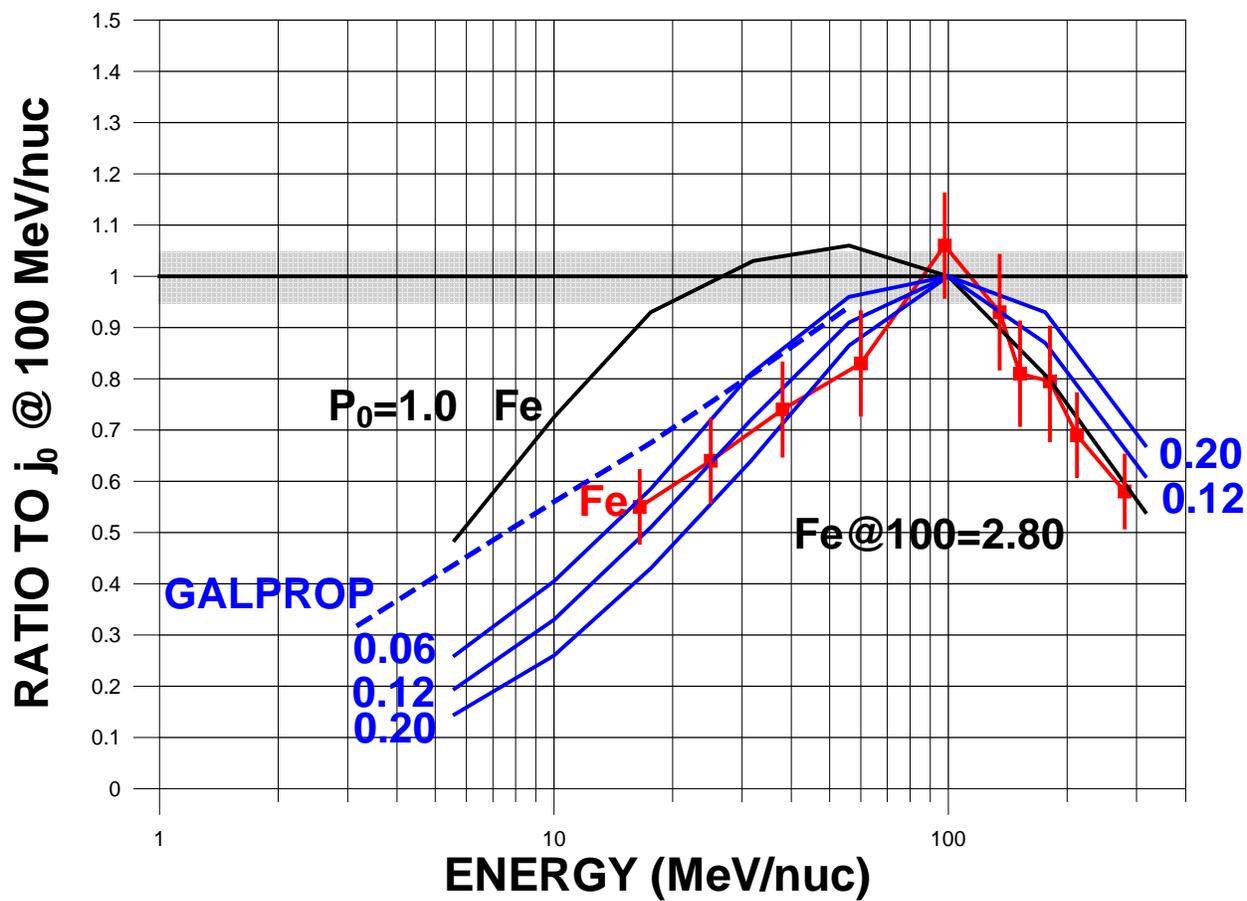

FIGURE 4

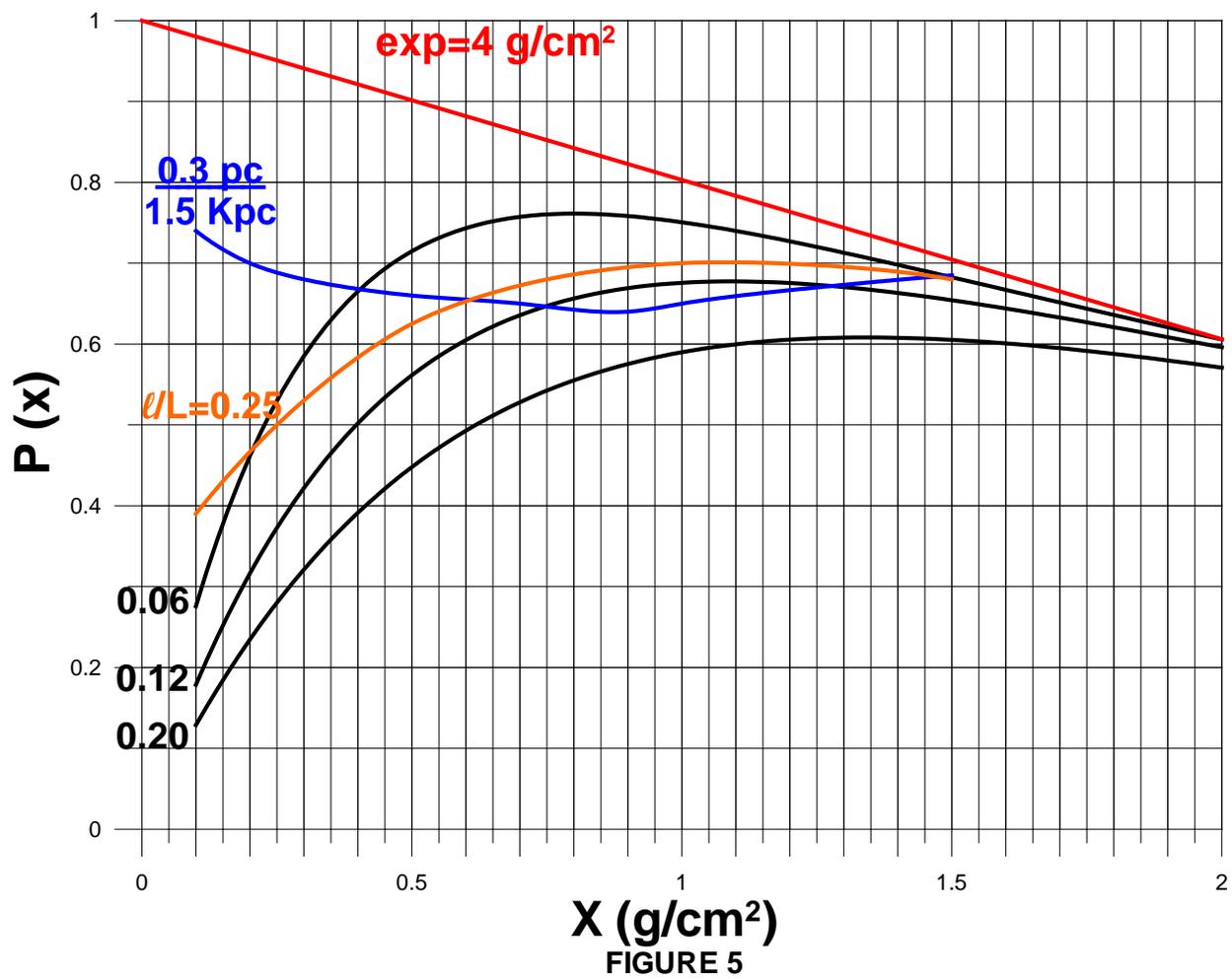

FIGURE 5